\documentclass[letter,twocolumn]{jpsj3}
\usepackage{graphicx}
\usepackage{txfonts}
\usepackage{bm}
\usepackage{amsmath}
\usepackage{color}

\def\gsim{\mathop {\vtop {\ialign {##\crcr 
$\hfil \displaystyle {>}\hfil $\crcr \noalign {\kern1pt \nointerlineskip } 
$\,\sim$ \crcr \noalign {\kern1pt}}}}\limits}
\def\lsim{\mathop {\vtop {\ialign {##\crcr 
$\hfil \displaystyle {<}\hfil $\crcr \noalign {\kern1pt \nointerlineskip } 
$\,\,\sim$ \crcr \noalign {\kern1pt}}}}\limits}

\title{Ultrasound Response in Quantum Critical $\beta$-YbAlB$_4$ and $\alpha$-YbAl$_{0.986}$Fe$_{0.014}$B$_4$}

\author{Shinji Watanabe}
\inst{Department of Basic Sciences, Kyushu Institute of Technology, Kitakyushu, Fukuoka 804-8550, Japan 
} 

\abst{
We analyze the key origin of quantum valence criticality in the heavy electron metal $\beta$-YbAlB$_4$ evidenced in the sister compound $\alpha$-YbAl$_{0.986}$Fe$_{0.014}$B$_4$. 
By constructing a realistic 
canonical
 model for $\beta$-YbAlB$_4$, we evaluate Coulomb repulsion between the 4f and 5d electrons at Yb $U_{\rm fd}\approx 6.2$~eV realizing the quantum critical point (QCP) of the Yb-valence transition. 
To reveal the Yb 5d contribution to the quantum critical state, we propose ultrasound measurement. 
We find that softening of elastic constants of not only the bulk modulus but also the shear moduli is caused by 
electric quadrupole fluctuations enhanced by critical 4f and 5d charge fluctuations for low temperatures at the valence QCP. 
Possible relevance of these results to $\beta$-YbAlB$_4$ and also $\alpha$-YbAl$_{1-x}$Fe$_x$B$_4$ is discussed.
}

\begin{document}
\maketitle

Quantum critical phenomena in strongly correlated metals have attracted great interest in condensed matter physics. 
In heavy-electron metal $\beta$-YbAlB$_4$ with an intermediate valence of Yb~\cite{Okawa2010}, unconventional quantum criticality as the susceptibility $\chi\sim T^{-0.5}$, the specific-heat coefficient $C/T\sim -\log T$, 
and the resistivity $\rho\sim T^{1.5} (\sim T)$ for $T\lsim 1$~K $(T\gsim 1~{\rm K})$ was observed~\cite{Nakatsuji}. 
Furthermore, a new type of scaling called $T/B$ scaling, 
where the magnetic susceptibility is expressed by a single scaling function of the ratio of temperature $T$ and magnetic field $B$, was observed in $\beta$-YbAlB$_4$~\cite{Matsumoto}. 
These phenomena have been shown to be explained by the theory of critical Yb-valence fluctuations in a unified way~\cite{WM2010,WM2014}. 
Recently, the evidence of the quantum critical point (QCP) of the Yb-valence transition giving rise to the quantum valence criticality as well as the $T/B$ scaling as observed in $\beta$-YbAlB$_4$ has been discovered experimentally in $\alpha$-YbAl$_{1-x}$Fe$_x$B$_4$ $(x=0.014)$~\cite{Kuga2018}. 

The key origin of the quantum valence criticality and the $T/B$ scaling is Coulomb repulsion $U_{\rm fd}$ between 4f and 5d electrons at Yb~\cite{WM2010,WM2014}.
Furthermore, novel odd-parity multipole degrees of freedom have recently been shown to be active by Yb 5d electrons theoretically~\cite{WM2019}. 
Hence, it is important to clarify the value of $U_{\rm fd}$ as well as to identify the Yb 5d contribution to the quantum critical state. 
In this Letter, we evaluate $U_{\rm fd}$ by constructing a realistic 
canonical
 model for $\beta$-YbAlB$_4$ and propose elastic-constant measurement to detect the Yb 5d contribution. 
We will show that not only the bulk modulus but also the shear moduli exhibit softening for low temperatures at the valence QCP. 
So far, ultrasound measurement in the unconventional quantum-critical materials has not been reported~\cite{Luthi}. 
Hence, the present study will pioneer this field.

\begin{figure}
\includegraphics[width=7cm]{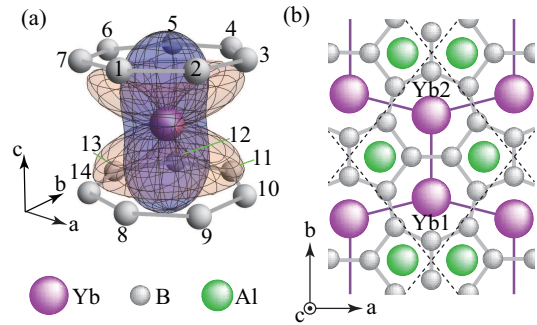}%
\caption{(color online) (a)Yb surrounded by 7 B rings. 
The squares of absolute values of spherical parts of the 4f wavefunction $\Psi^{4{\rm f}}_{\pm}(\hat{\bm r})=\langle\hat{\bm r}|\Psi^{4{\rm f}}_{\pm}\rangle$ (orange) and 5d wave function
$\Psi^{5{\rm d}}_{\pm}(\hat{\bm r})=\langle\hat{\bm r}|\Psi^{5{\rm d}}_{\pm}\rangle$ (purple) with $a_{5{\rm d}}=\sqrt{0.9}$ and $b_{5{\rm d}}=\sqrt{0.1}$ in Eq.~(\ref{eq:5dWF}) at Yb are shown (see text). 
(b) Top view of the lattice structure of $\beta$-YbAlB$_4$. Unit cell is the enclosed area by dashed lines.}
\label{fig:Yb_B}
\end{figure}

Let us start with the analysis of the crystalline-electric-field (CEF) in $\beta$-YbAlB$_4$ with orthorhombic crystal structure (No.65 $Cmmm$ $D_{2h}^{19}$)~\cite{Macaluso}. 
The CEF ground state of the 4f hole at Yb was theoretically proposed to be~\cite{NC2009} 
\begin{eqnarray}
|\Psi^{4{\rm f}}_{\pm}\rangle=|J=7/2, J_z=\pm 5/2\rangle, 
\label{eq:4fWF}
\end{eqnarray}
which has recently been supported by the linear polarization dependence of angle-resolved core level photoemission spectroscopy~\cite{Kuga2019}.
The conical wave function spreads toward 7 B rings in the upper and lower planes, which acquire the largest hybridization [see Fig.~\ref{fig:Yb_B}(a)]. 
As for the first-excited state, the mixture of $|7/2, \pm 1/2\rangle$ and $|7/2, \mp 3/2\rangle$, i.e. $|7/2, \pm1/2\rangle+\gamma|7/2, \mp3/2\rangle$, was shown to explain the 
anisotropic temperature dependence of the magnetic susceptibility, which earns the second-largest hybridization. 
This mixture is considered to be due to crystal fields of Al atoms that break the sevenfold symmetry of B rings [see Fig.~\ref{fig:Yb_B}(b)]~\cite{NC2009}.

As for the 5d state in Yb, the Hund's rule tells us that the $J=3/2$ state gives the lowest energy. 
The $\langle\hat{\bm r}|3/2, \pm1/2\rangle$ state is aligned along the $c$ direction while the $\langle\hat{\bm r}|3/2, \mp3/2\rangle$ state is lying in the $ab$ plane. 
The Yb 5d ground state is expected to be the mixture of both the states similarly to the first-excited Yb 4f state as
\begin{eqnarray}
|\Psi^{5{\rm d}}_{\pm}\rangle=a_{5{\rm d}}|3/2, \pm 1/2\rangle+b_{5{\rm d}}|3/2, \mp 3/2\rangle
\label{eq:5dWF}
\end{eqnarray}
%
with $a_{5{\rm d}}^2+b_{5{\rm d}}^2=1$. 
We expect that the wave function $\langle\hat{\bm r}|3/2, \pm1/2\rangle$ has larger hybridizations with 2p wave functions in the upper and lower B rings [see Fig.~\ref{fig:Yb_B}(a)] 
so that $|b_{5{\rm d}}/a_{5{\rm d}}|$ is considered to be small from the viewpoint of the hybridization picture of the CEF.

Next let us construct the effective Hamiltonian for the low-energy electronic states in $\beta$-YbAlB$_4$. 
Since now we are interested in the properties of the ground state as well as the low temperature much smaller than the first excited CEF energy $(\Delta\approx 80~{\rm K})$~\cite{NC2009}, 
we consider the following model in the hole picture which consists of the 4f state $|\Psi_{\pm}^{4{\rm f}}\rangle$ and 5d state $|\Psi_{\pm}^{5{\rm d}}\rangle$ at Yb as well as the $2p$ state at B. 
\begin{eqnarray}
H=\sum_{i}\sum_{\alpha=1,2}\left[
H^{\rm f}_{i\alpha}+H^{\rm pf}_{i\alpha}+H^{\rm pd}_{i\alpha}+H^{U_{\rm fd}}_{i\alpha}
\right]+H^{\rm d}+H^{\rm p}, 
\label{eq:Hamil}
\end{eqnarray}
where $i=1, \cdots N$ with $N$ being the number of the unit cell and 
$\alpha=1, 2$ denote the Yb1 and Yb2 sites respectively [see Fig.~\ref{fig:Yb_B}(b)]. 

The 4f part is 
\begin{eqnarray}
H^{\rm f}_{i\alpha}=\varepsilon_{\rm f}\sum_{\eta=\pm}n^{\rm f}_{i\alpha \eta}+Un^{\rm f}_{i\alpha+}n^{\rm f}_{i\alpha-} 
\label{eq:H_4f}
\end{eqnarray}
with $n^{\rm f}_{i\alpha\eta}\equiv f^{\dagger}_{i\alpha\eta}f_{i\alpha\eta}$, where the $f^{\dagger}_{i\alpha\eta}$ $(f_{i\alpha\eta})$ operators create (annifilate) 4f holes with the Kramers state 
$\eta=\pm$ of $|\Psi_{\pm}^{4{\rm f}}\rangle$ in Eq.~(\ref{eq:4fWF}) 
and $\varepsilon_{\rm f}$ is the energy level. 
Here $U$ is the onsite Coulomb repulsion. 

The 2p part is
\begin{eqnarray}
H^{\rm p}=\sum_{\langle j,j'\rangle}\sum_{\sigma=\uparrow,\downarrow}\sum_{m, m'=z, \pm}t^{\rm pp}_{jm,j'm'}p^{\dagger}_{jm\sigma}p_{j'm'\sigma},
\label{eq:H_p}
\end{eqnarray}
where $\langle j,j'\rangle$ denotes the nearest neighbor (N.N.) B sites in the $ab$ plane and $c$ direction [see Figs.~\ref{fig:Yb_B}(a) and \ref{fig:Yb_B}(b)] and $t^{\rm pp}_{jm,j'm'}$ is the transfer integral between the 2p states. 
Here, the $p^{\dagger}_{jm\sigma}$ $(p_{jm\sigma})$ operators create (annifilate) 2p holes 
at the $j$-th B site
with $m=p_z, p_{\pm}$, where $p_{\pm}$ is defined by $p_{\pm}\equiv(p_x \pm 
i
p_y)/\sqrt{2}$, and spin $\sigma=\uparrow, \downarrow$. 
For simplicity, the energy levels of the 2p state at each B site are set to be the same, which is taken as the origin of the energy.

The 4f-2p hybridization is 
\begin{eqnarray}
H^{\rm pf}_{i\alpha}=\sum_{\langle i\alpha, j\rangle}\sum_{m, \sigma, \eta}(V^{\rm pf}_{jm\sigma, i\alpha \eta}p^{\dagger}_{jm\sigma}f_{i\alpha \eta}+h.c.),
\label{eq:H_pf}
\end{eqnarray}
where $\langle i\alpha, j\rangle$ denotes the N.N. pair of the Yb$\alpha$ site in the $i$-th unit cell and the $j$-th B site with $j=1-7$ (upper plane) 
and $j=8-14$ (lower plane) in Fig.~\ref{fig:Yb_B}(a). 

The 5d part is
\begin{eqnarray}
H^{\rm d}=\varepsilon_{\rm d}\sum_{i}\sum_{\alpha=1,2}\sum_{\eta=\pm}n^{\rm d}_{i\alpha\eta}
+\sum_{\langle i\alpha, i'\alpha'\rangle}\sum_{\eta, \eta'=\pm}t^{\rm dd}_{i\alpha\eta, i'\alpha'\eta'}d^{\dagger}_{i\alpha\eta}d_{i'\alpha'\eta'} 
\label{eq:H_d}
\end{eqnarray}
with $n^{\rm d}_{i\alpha\eta}\equiv d^{\dagger}_{i\alpha\eta}d_{i\alpha\eta}$, where the $d^{\dagger}_{i\alpha\eta}$ $(d_{i\alpha\eta})$ operators create (annihilate) 5d holes with the Kramers state $\eta=\pm$ of $|\Psi_{\pm}^{5{\rm d}}\rangle$ in Eq.~(\ref{eq:5dWF})
and $\varepsilon_{\rm d}$ is the energy level. 
Here $t^{\rm dd}_{i\alpha\eta, i'\alpha'\eta'}$ is the transfer integral between the 5d states and $\langle i\alpha, i'\alpha'\rangle$ denotes the pair of the Yb sites for the N.N. (in the $c$ direction), the second N.N. (Yb1-Yb2 sites inside the unit cell), and the third N.N. (Yb1-Yb2 sites between the adjacent unit cells) [see Fig.~\ref{fig:Yb_B}(b)].

The 5d-2p hybridization is 
\begin{eqnarray}
H^{\rm pd}_{i\alpha}=\sum_{\langle i\alpha, j\rangle}\sum_{m, \sigma, \eta}(V^{\rm pd}_{jm\sigma, i\alpha \eta}p^{\dagger}_{jm\sigma}d_{i\alpha \eta}+h.c.). 
\label{eq:H_pf}
\end{eqnarray}

The 4f-5d Coulomb repulsion at Yb is
\begin{eqnarray}
H^{U_{\rm fd}}_{i\alpha}=U_{\rm fd}\sum_{\eta=\pm}\sum_{\eta'=\pm}n^{\rm f}_{i
\alpha
\eta}n^{\rm d}_{i
\alpha
\eta'}. 
\label{eq:H_Ufd}
\end{eqnarray}
%

We note that onsite 4f-5d hybridization occurs at Yb via the 4f-2p and 2p-5d hybridizations in Eq.~(\ref{eq:Hamil}), as shown in Ref.~\citen{WM2019}. This is nothing but the odd-parity CEF due to the local violation of the inversion symmetry at the Yb site by the sevenfold configuration of the surrounding B sites~\cite{WM2019}. The 4f-5d hybridization between the Yb sites is ignored since its magnitude is negligibly small. 
The other interactions such as 5d-5d Coulomb repulsion are ignored since their effects are regarded to be renormalized into the conduction bands in Eq.~(\ref{eq:Hamil}). 

To analyze electronic states in $\beta$-YbAlB$_4$, 
we apply the slave-boson mean-field (MF) theory to Eq.~(\ref{eq:Hamil})~\cite{Read1983}. 
To describe the state for $U=\infty$ causative of heavy electrons, we consider $f^{\dagger}_{i\alpha\eta}b_{i\alpha}$ instead of $f^{\dagger}_{i\alpha\eta}$ in Eq.~(\ref{eq:Hamil}) 
by introducing the slave-boson operator $b_{i\alpha}$ to describe the $f^0$-hole state and require the constraint $\sum_{i\alpha}\lambda_{i\alpha}(\sum_{\eta=\pm}n^{\rm f}_{i\alpha\eta}+b^{\dagger}_{i\alpha}b_{i\alpha}-1)$. 
Here $\lambda_{i\alpha}$ is the Lagrange multiplier. 
To $H^{U_{\rm fd}}_{i\alpha}$ in Eq.~(\ref{eq:H_Ufd}), we apply the MF decoupling as 
$U_{\rm fd}n^{\rm f}_{i\alpha\eta}n^{\rm d}_{i\alpha\eta'}\approx U_{\rm fd}\bar{n}^{\rm f}_{\alpha}n^{\rm d}_{i\alpha\eta'}+\bar{R}_{\alpha}n^{\rm f}_{i\alpha\eta}-\bar{R}_{\alpha}\bar{n}^{\rm f}_{\alpha}$, 
with $\bar{R}_{\alpha}\equiv U_{\rm fd}\bar{n}^{\rm d}_{\alpha}$
and $\bar{n}^{\rm f (d)}_{\alpha}\equiv\sum_{i\eta}\langle n^{\rm f (d)}_{i\alpha\eta}\rangle/N$. 
Since we focus on the paramagnetic-metal phase, it is natural to 
approximate the MFs to uniform ones $\bar{b}_{\alpha}=\langle b_{i\alpha}\rangle$ and $\bar{\lambda}_{\alpha}=\lambda_{i\alpha}$. 
Then, by optimizing the Hamiltonian as $\partial\langle H\rangle/\partial\bar{b}_{\alpha}=0$, $\partial\langle H\rangle/\partial\bar{\lambda}_{\alpha}=0$,
and $\partial\langle H\rangle/\partial\bar{R}_{\alpha}=0$, 
we obtain the set of the MF equations : 
%
$\frac{1}{N}\sum_{{\bm k}\eta}\langle f_{{\bm k}\alpha \eta}^{\dagger}f_{{\bm k}\alpha \eta}\rangle+\bar{b}_{\alpha}^2=1$, 
$\frac{1}{2N}\sum_{{\bm k}}\left[\sum_{\eta\xi m\sigma}V^{\rm pf*}_{{\bm k},\xi m\sigma, \alpha{\eta}}\langle f^{\dagger}_{{\bm k}\alpha{\eta}}p_{{\bm k}\xi m\sigma}\rangle
+h.c.\right]+\bar{\lambda}_{\alpha}\bar{b}_{\alpha}=0$, 
and 
$\bar{n}^{\rm f}_{\alpha}=\frac{1}{N}\sum_{{\bm k}\eta}\langle f^{\dagger}_{{\bm k}\alpha{\eta}}f_{{\bm k}\alpha{\eta}}\rangle$.
%
Here, $\xi$ specifies the N.N. B sites for the Yb$\alpha$ site [see Fig.~\ref{fig:Yb_B}(b)]. 
We solve the MF equations together with the equation for the filling $\bar{n}\equiv\sum_{\alpha=1,2}(\bar{n}^{\rm f}_{\alpha}+\bar{n}^{\rm d}_{\alpha})/4+\sum_{j=1}^{8}\bar{n}^{p}_{j}/16$ 
with $\bar{n}^{p}_{j}\equiv\sum_{{\bm k}m\sigma}\langle p^{\dagger}_{{\bm k}jm\sigma}p_{{\bm k}jm\sigma}\rangle/(3N)$ self-consistently. 

In the calculation of $t^{\rm pp}_{jm,j'm'}$, $t^{\rm dd}_{i\alpha\eta, i'\alpha'\eta'}$, $V^{\rm pf}_{jm\sigma, i\alpha \eta}$, and $V^{\rm pd}_{jm\sigma, i\alpha \eta}$, 
we need to input the Slater-Koster parameters~\cite{Slater,Takegahara}. 
Following the argument of the linear muffin-tin orbital 
(LMTO)
method~\cite{Andersen1977}, we employ the general relation
$(pp\pi)=-(pp\sigma)/2$, $(pd\pi)=-(pd\sigma)/\sqrt{3}$, $(pf\pi)=-(pf\sigma)/\sqrt{3}$, $(dd\pi)=-2(dd\sigma)/3$, and $(dd\delta)=(dd\sigma)/6$. 
In the hole picture, we take the energy unit as $(pp\sigma)=-1.0$ and set 
$(pd\sigma)=0.6$, $(pf\sigma)=-0.3$, and $(dd\sigma)=0.4$ as typical values. 
We note that distance dependences of transfer integrals and hybridizations between the $l$ and $l'$ states with $l$ being the orbital angular momentum are set so as to follow $\sim 1/r^{l+l'+1}$ with $r$ being the distance between the two atoms in the LMTO method~\cite{Andersen1977}.
We set $a_{5{\rm d}}=\sqrt{0.9}$ and $b_{5{\rm d}}=\sqrt{0.1}$ in Eq.~(\ref{eq:5dWF}) as the representative case [see Fig.~\ref{fig:Yb_B}(a)].
The calculated band structure near the Fermi level for $\varepsilon_{\rm d}=-1$ and $\varepsilon_{\rm f}\approx -2.1$ at $\bar{n}=1$ 
well reproduces the recent photoemission data~\cite{Cedric} and then we adopt these parameters in this study. 
We performed the numerical calculations in the $N=8^3$, $16^3$, and $32^3$ systems and will show the results in $N=32^3$. 

\begin{figure}[t]
\includegraphics[width=7cm]{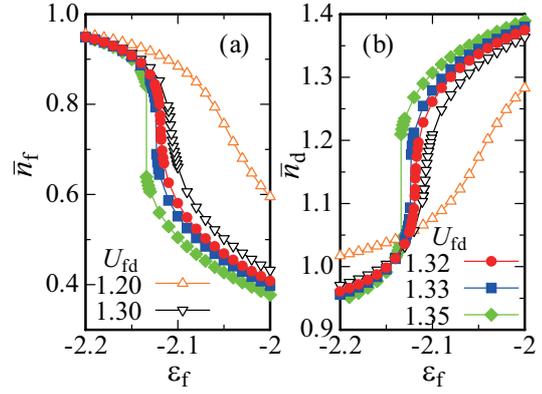}
\caption{(color online) The $\varepsilon_{\rm f}$ dependences of (a) the 4f-hole number and (b) the 5d-hole number at Yb 
for $U_{\rm fd}=1.20$ (open triangle), $1.30$ (open inverted triangle), $1.32$ (filled circle), $1.33$ (filled square), and $1.35$ (filled diamond).}
\label{fig:nf_nd}
\end{figure}

The $\varepsilon_{\rm f}$ dependences of the 4f-hole number $\bar{n}^{\rm f} (=\bar{n}^{\rm f}_1=\bar{n}^{\rm f}_2)$ and the 5d-hole number $\bar{n}^{\rm d} (=\bar{n}^{\rm d}_1=\bar{n}^{\rm d}_2)$ at Yb for the ground state is shown in Figs.~\ref{fig:nf_nd}(a) and \ref{fig:nf_nd}(b), respectively. 
As $\varepsilon_{\rm f}$ increases, $\bar{n}_{\rm f}$ $(\bar{n}_{\rm d})$ decreases (increases). 
As $U_{\rm fd}$ increases, the $\bar{n}_{\rm f}$ $(\bar{n}_{\rm d})$ change becomes sharp and for $U_{\rm fd}=1.32$ the slope $-\partial\bar{n}_{\rm f}/\partial\varepsilon_{\rm f}$ and $\partial\bar{n}_{\rm d}/\partial\varepsilon_{\rm f}$ diverges 
at $\varepsilon_{\rm f}=-2.1185$. 
For $U_{\rm fd}>1.32$, a jump in $\bar{n}_{\rm f}$ and $\bar{n}_{\rm d}$ appears as shown in Figs.~\ref{fig:nf_nd}(a) and \ref{fig:nf_nd}(b), respectively, indicating the first-order valence transition. 
From these results, the QCP of the valence transition is identified to be $(\varepsilon_{\rm f}^{\rm QCP}, U_{\rm fd}^{\rm QCP})=(-2.1185, 1.32)$. 
We note that $\bar{n}_{\rm f}=0.74$ realized at the QCP, which is favorably compared with Yb$^{+2.75}$ observed in $\beta$-YbAlB$_4$ at $T=20$~K~\cite{Okawa2010}. 
If we estimate $(pp\sigma)\approx 4.7$~eV from the first-principles calculation in B~\cite{Papa}, $U_{\rm fd}^{\rm QCP}$ is evaluated to be $U_{\rm fd}^{\rm QCP}\approx 6.2$~eV. 
This value seems reasonable since $U_{\rm fd}$ is onsite Coulomb repulsion at Yb. 
The examination of the $U_{\rm fd}$ value by direct measurements such as the partial-fluorescence-yield method of X ray~\cite{Tonai2017} in $\beta$-YbAlB$_4$ and 
$\alpha$-YbAl$_{1-x}$Fe$_x$B$_4$ $(x=0.014)$ is highly desirable.

Next, let us proceed to the framework beyond the MF theory. 
Namely, we calculate the susceptibility by the random phase approximation (RPA) with respect to $U_{\rm fd}$ as the corrections for the MF state. 
This enables us to analyze the irreducible susceptibility systematically. 
The RPA susceptibility is calculated as~\cite{THU2004} 
\begin{eqnarray}
& &\chi^{\alpha\beta}_{\ell_1\eta_1\ell_2\eta_2\ell_3\eta_3\ell_4\eta_4}({\bm q},\omega)
=\bar{\chi}^{\alpha\beta}_{\ell_1\eta_1\ell_2\eta_2\ell_3\eta_3\ell_4\eta_4}({\bm q},\omega)
\nonumber
\\
&-&\sum_{\gamma}\sum_{\tau\tau'}\sum_{m\ne m'}
\bar{\chi}^{\alpha\gamma}_{\ell_1\eta_1\ell_2\eta_2m\tau m\tau}({\bm q},\omega)
U_{\rm fd}
\chi^{\gamma\beta}_{m'\tau' m'\tau'\ell_3\eta_3\ell_4\eta_4}({\bm q},\omega)
\nonumber
\\
&+&\sum_{\gamma}\sum_{\tau\tau'}\sum_{m\ne m'}\bar{\chi}^{\alpha\gamma}_{\ell_1\eta_1\ell_2\eta_2 m\tau m'\tau'}({\bm q},\omega)
U_{\rm fd}
\chi^{\gamma\beta}_{m\tau m'\tau'\ell_3\eta_3\ell_4\eta_4}({\bm q},\omega),
\nonumber
\\
& &
\label{eq:RPA}
\end{eqnarray}
where the susceptibility is defined by 
\begin{eqnarray}
\chi^{\alpha\beta}_{\ell_1\eta_1\ell_2\eta_2\ell_3\eta_3\ell_4\eta_4}({\bm q},\omega)
\equiv\frac{i}{N}\sum_{{\bm k}{\bm k'}}\int_{0}^{\infty}dte^{i\omega t}
\nonumber
\\
\times
\langle[c^{\dagger}_{{\bm k}\alpha\ell_1\eta_1}(t)c_{{\bm k}+{\bm q}\alpha\ell_2\eta_2}(t), c^{\dagger}_{{\bm k'}+{\bm q}\beta\ell_4\eta_4}c_{{\bm k'}\beta\ell_3\eta_3}]\rangle. 
\label{eq:chi_def}
\end{eqnarray}
Here, $\ell=1 (2)$ denotes the f (d) orbital and 
$\bar{\chi}^{\alpha\gamma}_{\ell_1\eta_1\ell_2\eta_2\ell_3\eta_3 \ell_4\eta_4}({\bm q},\omega)$ represents the susceptibility calculated for the MF state.
The valence transition is caused by the inter-orbital Coulomb repulsion $U_{\rm fd}$ after the formation of heavy quasiparticles by $U\to\infty$. 
We first obtained the QCP of the valence transition within the MF theory for $U\to\infty$ and then to take into account further critical fluctuations caused by $U_{\rm fd}$, 
we employ Eq.~(\ref{eq:RPA}). 
The RPA susceptibility in Eq.~(\ref{eq:RPA}) is expressed by the 
$32\times 32$ matrix $\hat{\chi}$, $\hat{\bar{\chi}}$ and $\hat{U}$ in the symmetrized form as 
\begin{eqnarray}
\hat{\chi}&=&\hat{\bar{\chi}}+\hat{\bar{\chi}}\hat{U}\hat{\bar{\chi}}+\hat{\bar{\chi}}\hat{U}\hat{\bar{\chi}}\hat{U}\hat{\bar{\chi}}+\cdots, 
\nonumber
\\
&=&\hat{\bar{\chi}}^{1/2}\left(\hat{1}-\hat{\bar{\chi}}^{1/2}\hat{U}\hat{\bar{\chi}}^{1/2}\right)^{-1}\hat{\bar{\chi}}^{1/2},
\label{eq:RPA_chi}
\end{eqnarray}
where $\hat{\bar{\chi}}^{1/2}$ is the matrix satisfying $\hat{\bar{\chi}}=\hat{\bar{\chi}}^{1/2}\hat{\bar{\chi}}^{1/2}$ and $\hat{1}$ is the identity matrix.
The interaction matrix $\hat{U}$ has the elements of $U_{\rm fd}$ for $\ell_1=\ell_3\ne\ell_2=\ell_4$, $\eta_1=\eta_3$, and $\eta_2=\eta_4$
and $-U_{\rm fd}$ for $\ell_1=\ell_2\ne\ell_3=\ell_4$, $\eta_1=\eta_2$, and $\eta_3=\eta_4$. 

The critical point in this RPA formalism is identified by 
\begin{eqnarray}
{\rm det}\left(\hat{1}-\hat{\bar{\chi}}^{1/2}\hat{U}\hat{\bar{\chi}}^{1/2}\right)=0. 
\label{eq:RPA_det0}
\end{eqnarray}
By using the MF states obtained in the calculation in Fig.~\ref{fig:nf_nd}, we calculate $\hat{\bar{\chi}}$ by Eq.~(\ref{eq:chi_def}) and solve Eq.~(\ref{eq:RPA_det0}). 
Then, critical point in this formalism is identified to be $(\varepsilon_{\rm f}^{\rm c RPA}, U_{\rm fd}^{\rm c RPA})=(-2.1185, 0.4788)$.

In the present multi-orbital system, there exist total charge fluctuation and relative charge fluctuation with respect to the 4f and 5d orbitals, 
which are defined by 
\begin{eqnarray}
\chi^{\alpha\beta}_{n_{\rm f}\pm n_{\rm d}}({\bm q},\omega)&=&\frac{i}{N}\int_{0}^{\infty}dte^{i\omega t}
\nonumber
\\
& &
\times
\langle[\delta n^{\rm f}_{\bm{q}\alpha}(t)\pm \delta n^{\rm d}_{\bm{q}\alpha}(t), \delta n^{\rm f}_{-\bm{q}\beta}(0)\pm \delta n^{\rm d}_{-\bm{q}\beta}(0) ]\rangle, 
\label{eq:chi_nf_nd1}
\\
&=&\sum_{\eta\eta'}\left[
\sum_{\ell}\chi^{\alpha\beta}_{\ell\eta \ell\eta \ell\eta' \ell\eta'}({\bm q}, \omega) 
\pm\sum_{\ell\ne\ell'}\chi^{\alpha\beta}_{\ell\eta \ell\eta \ell'\eta' \ell'\eta'}({\bm q}, \omega)
\right] 
\nonumber
\\
& &
\label{eq:chi_nf_nd}
\end{eqnarray}
with $+(-)$ denoting the total (relative) charge susceptibility. Here $\delta\hat{\cal O}$ is defined as $\delta\hat{\cal O}\equiv\hat{\cal O}-\langle\hat{\cal O}\rangle$ 
and $n^{\rm f (d)}_{{\bm q}\alpha}$ is given by
%
$n^{\rm f (d)}_{{\bm q}\alpha}=\sum_{i}e^{-i{\bm q}\cdot\bm{r}_i}n^{\rm f (d)}_{i\alpha}$.
%

\begin{figure}[t]
\includegraphics[width=7cm]{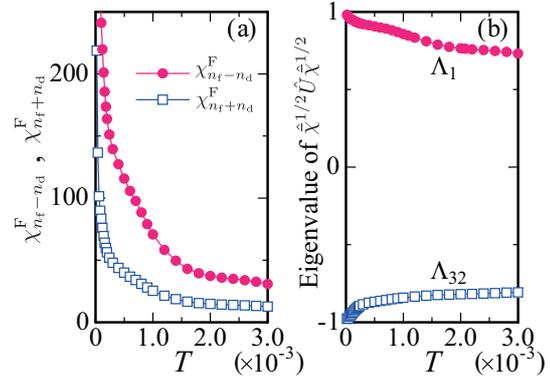}
\caption{(color online)
Temperature dependence of (a) relative and total charge susceptibilities and (b) eigenvalues of $\hat{\bar{\chi}}^{1/2}\hat{U}\hat{\bar{\chi}}^{1/2}$ 
for $(\varepsilon_{\rm f}^{\rm cRPA}, U_{\rm fd}^{\rm cRPA})$. 
}
\label{fig:nf_nt_nr}
\end{figure}

In Fig.~\ref{fig:nf_nt_nr}(a), we plot the temperature dependence of the uniform relative-charge fluctuation between 4f and 5d holes
%
$\chi^{\rm F}_{n_{\rm f}-n_{\rm d}}=\lim_{{\bm q}\to{\bm 0}}\chi^{\rm F}_{n_{\rm f}-n_{\rm d}}({\bm q}, \omega=0)$
%
and the uniform total-charge fluctuation between 4f and 5d holes
%
$\chi^{\rm F}_{n_{\rm f}+n_{\rm d}}=\lim_{{\bm q}\to{\bm 0}}\chi^{\rm F}_{n_{\rm f}+n_{\rm d}}({\bm q}, \omega=0)$ 
%
for $(\varepsilon_{\rm f}^{\rm c RPA}, U_{\rm fd}^{\rm c RPA})$. 
Here, $\chi^{\rm F}_{n_{\rm f}\pm n_{\rm d}}({\bm q}, \omega)$ is defined by 
%
$\chi^{\rm F}_{n_{\rm f}\pm n_{\rm d}}({\bm q}, \omega)=\sum_{\alpha\beta}\chi^{\alpha\beta}_{n_{\rm f}\pm n_{\rm d}}({\bm q},\omega).$ 
%

At the valence QCP, $\chi^{\rm F}_{n_{\rm f}-n_{\rm d}}$ diverges for $T\to 0$, while $\chi^{\rm F}_{n_{\rm f}+n_{\rm d}}$ 
remains finite for $T\to 0$ although $\chi^{\rm F}_{n_{\rm f}+n_{\rm d}}$ increases at low temperatures. 
This was confirmed by the temperature dependence of eigenvalues of $\hat{\bar{\chi}}^{1/2}\hat{U}\hat{\bar{\chi}}^{1/2}$ in Fig.3(b). 
The eigenvector analysis tells us that 
the maximum and minimum eigenvalues $\Lambda_1$, $\Lambda_{32}$ corresponds to the relative and total charge fluctuations $\chi^{\rm F}_{n_{\rm f}-n_{\rm d}}$, $\chi^{\rm F}_{n_{\rm f}+n_{\rm d}}$, respectively. 
Figure3(b) shows $\Lambda_1(T)\to 1$ for $T\to 0$, which satisfies Eq.~(\ref{eq:RPA_det0}) at $T=0$, indicating the QCP within the RPA.
The divergence of the relative-charge fluctuation is naturally understood from Figs.~\ref{fig:nf_nd}(a) and \ref{fig:nf_nd}(b).

Next let us discuss the electric quadrupole susceptibility 
\begin{eqnarray}
\chi^{\alpha\beta}_{O_{\Gamma}}({\bm q},\omega)=\frac{i}{N}\int_{0}^{\infty}dte^{i\omega t}\langle[\delta\hat{O}^{\Gamma}_{\bm{q}\alpha}(t), \delta\hat{O}^{\Gamma}_{\bm{q}\beta}(0)^{\dagger}]\rangle, 
\label{eq:chi_E4}
\end{eqnarray}
where $\hat{O}^{\Gamma}_{{\bm q}\alpha}$ is given by 
%
$\hat{O}^{\Gamma}_{{\bm q}\alpha}=\sum_{i}e^{-i{\bm q}\cdot\bm{r}_i}\hat{O}_{i\alpha}$ 
%
with the irreducible representation $\Gamma$.
Here, $\hat{O}^{\Gamma}_{i\alpha}$ is expressed as
\begin{eqnarray}
\hat{O}^{\Gamma}_{i\alpha}=\sum_{\ell\ell'}\sum_{\eta\eta'}O^{\Gamma}_{\alpha\ell\eta, \alpha\ell'\eta'}c^{\dagger}_{i\alpha\ell\eta}c_{i\alpha\ell'\eta'}, 
\label{eq:Oi}
\end{eqnarray}
where $O^{\Gamma}_{\alpha\ell\eta, \alpha\ell'\eta'}$ is the form factor given by 
%
$O^{\Gamma}_{\alpha\ell\eta, \alpha\ell'\eta'}=\langle\alpha\ell\eta|\hat{O}_{\Gamma}|\alpha\ell'\eta'\rangle$. 
%
Then, Eq. (\ref{eq:chi_E4}) leads to
\begin{eqnarray}
\chi^{\alpha\beta}_{O_{\Gamma}}({\bm q},\omega)&=&\sum_{\ell_1\ell_2\ell_3\ell_4}\sum_{\eta_1\eta_2\eta_3\eta_4}O^{\Gamma}_{\alpha\ell_1\eta_1, \alpha\ell_2\eta_2}\chi_{\ell_1\eta_1\ell_2\eta_2\ell_3\eta_3\ell_4\eta_4}^{\alpha\beta}({\bm q},\omega)
\nonumber
\\
& & \times O^{\Gamma}_{\beta\ell_4\eta_4, \beta\ell_3\eta_3}, 
\label{eq:chi_E4_2}
\end{eqnarray}

In $\beta$-YbAlB$_4$, the crystal structure is orthorhombic and the point group is $D_{2h}$. Then $\Gamma$ is given by $\Gamma=x^2$, $y^2$, $z^2$, $xy$, $xz$, and $yz$. 
Here, following Ref.~\citen{Luthi}, we consider the basis $x^2+y^2+z^2$, $2z^2-x^2-y^2$, and $x^2-y^2$ for the $A_{\rm 1g}$ symmetry instead of $x^2$, $y^2$, and $z^2$, which also belong to the same $A_{\rm 1g}$ representation.  
The corresponding operators of $\hat{O}_{\Gamma}$ are expressed as the symmetrized form of the operators of the total angular momentum as 
$\hat{O}_{x^2+y^2+z^2}=J_x^2+J_y^2+J_z^2$, $\hat{O}_{2z^2-x^2-y^2}=2J_z^2-J_x^2-J_y^2$, $\hat{O}_{x^2-y^2}=J_x^2-J_y^2$, $\hat{O}_{xy}=J_xJ_y+J_yJ_x$, $\hat{O}_{xz}=J_xJ_z+J_zJ_x$, and $\hat{O}_{yz}=J_yJ_z+J_zJ_y$. 
The form factors are calculated for the 4f state $|\Psi^{4{\rm f}}\rangle$ by using Eq.~(\ref{eq:4fWF}) as
%
$\langle\alpha{1}\pm|\hat{O}_{x^2+y^2+z^2}|\alpha 1\pm\rangle=63/4$, 
$\langle\alpha 1{\pm}|\hat{O}_{2z^2-x^2-y^2}|\alpha 1{\pm}\rangle=3$, 
%
and $\langle\alpha 1{\pm}|\hat{O}_{\Gamma}|\alpha 1{\pm}\rangle=0$ for $\Gamma=x^2-y^2, xy, yz$, and $zx$. 
The form factors for the 5d state $|\Psi^{5{\rm d}}\rangle$ are calculated by using Eq.~(\ref{eq:5dWF}) as 
%
$\langle\alpha 2{\pm}|\hat{O}_{x^2+y^2+z^2}|\alpha 2{\pm}\rangle=15/4$, 
$\langle\alpha 2{\pm}|\hat{O}_{2z^2-x^2-y^2}|\alpha 2{\pm}\rangle=-3a_{5{\rm d}}^2+3b_{5{\rm d}}^2$,
$\langle\alpha 2{\pm}|\hat{O}_{x^2-y^2}|\alpha 2{\pm}\rangle=2\sqrt{3}a_{5{\rm d}}b_{5{\rm d}}$,
%
and 
$\langle\alpha 2{\pm}|\hat{O}_{\Gamma}|\alpha 2{\pm}\rangle=0$ for $\Gamma=xy$, $yz$, and $zx$.
Then, we obtain 
\begin{eqnarray}
\chi^{\alpha\beta}_{O_{\Gamma}}({\bm q},\omega)=\sum_{\ell=1,2}\left[\sum_{\eta}O^{\Gamma}_{\alpha \ell\eta, \alpha \ell\eta}\chi^{\alpha\beta}_{\ell\eta \ell\eta \ell\eta \ell\eta}({\bm q},\omega)O^{\Gamma}_{\beta \ell\eta, \beta \ell\eta}
\right.
\nonumber
\\
+
\left.
\sum_{\eta}O^{\Gamma}_{\alpha \ell\eta, \alpha \ell\eta}\chi^{\alpha\beta}_{\ell\eta \ell\eta \ell-\eta \ell-\eta}({\bm q},\omega)O^{\Gamma}_{\beta \ell-\eta, \beta \ell-\eta}
\right]
\label{eq:chi_E4_3}
\end{eqnarray}
for $\Gamma=x^2+y^2+z^2$ and $2z^2-x^2-y^2$ and obtain 
$\chi_{O_{x^2-y^2}}^{\alpha\beta}({\bm q},\omega)$ by setting $\ell=2$ in the right hand side (r.h.s.) of Eq.~(\ref{eq:chi_E4_3}).

The elastic constant is expressed as
\begin{eqnarray}
C_{\Gamma}=C_{\Gamma}^{(0)}-g_{\Gamma}^2\chi_{\Gamma}, 
\label{eq:elastic_C}
\end{eqnarray}
where $C_{\Gamma}^{(0)}$ is the elastic constant of the background and $g_{\Gamma}$ is the quadrupole-strain coupling constant. 
Here, $\chi_{\Gamma}$ is defined by 
$\chi_{\Gamma}=\lim_{{\bm q}\to{\bm 0}}\chi_{\Gamma}^{\rm F}({\bm q},\omega=0)$ 
with $\chi^{\rm F}_{O_{\Gamma}}({\bm q},\omega)=\sum_{\alpha\beta}\chi^{\alpha\beta}_{O_{\Gamma}}({\bm q},\omega)$.

\begin{figure}[t]
\includegraphics[width=7cm]{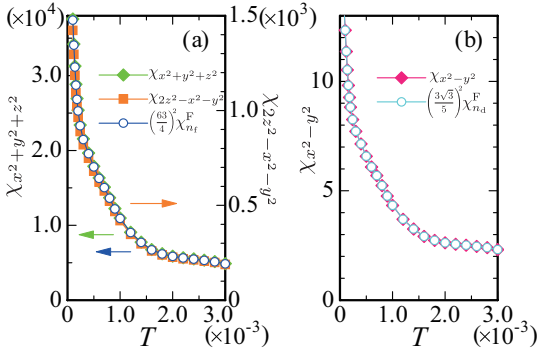}
\caption{(color online)
Temperature dependence of (a) electric quadrupole susceptibilities $\chi_{x^2+y^2+z^2}$ and $\chi_{2z^2-x^2-y^2}$ and charge susceptibility $(63/4)^2\chi_{n_{\rm f}}^{\rm F}$
and (b) electric quadrupole susceptibility $\chi_{x^2-y^2}$ and charge susceptibility $(3\sqrt{3}/5)^2\chi_{n_{\rm d}}^{\rm F}$ 
for $(\varepsilon_{\rm f}^{\rm cRPA}, U_{\rm fd}^{\rm cRPA})$. 
}
\label{fig:chi_E4}
\end{figure}

In Fig.4(a), we plot the temperature dependence of $\chi_{x^2+y^2+z^2}$ and $\chi_{2z^2-x^2-y^2}$ for $(\varepsilon_{\rm f}, U_{\rm fd})=(\varepsilon_{\rm f}^{\rm cRPA}, U_{\rm fd}^{\rm cRPA})$. 
As $T$ decreases, $\chi_{x^2+y^2+z^2}$ and $\chi_{2z^2-x^2-y^2}$ increase. 
For $T<10^{-3}$, i.e., in the low-temperature region below $55~{\rm K}<\Delta/k_{\rm B}$ with $k_{\rm B}$ being the Boltzmann constant, a remarkable enhancement emerges  
 [here, $T=10^{-3}\sim 55$ K is estimated by assuming $(pp\sigma)\approx 4.7$~eV as above]. 
Both show the same temperature dependence although the former is one order of magnitude larger than the latter reflecting the difference of the form factors as shown above. 
We also plot $(63/4)^2\chi^{\rm F}_{n_{\rm f}}$ with $\chi^{\rm F}_{n_{\rm f}}=\lim_{{\bm q}\to 0}\chi^{\rm F}_{n_{\rm f}}({\bm q}, \omega=0)$ in Fig.4(a). 
Here, $\chi_{n_{\rm f (d)}}^{\rm F}$ is obtained by setting the d (f) part zero in Eq.~(\ref{eq:chi_nf_nd1}), which is expressed by 
the first term in the r.h.s. of Eq.~(\ref{eq:chi_nf_nd}) with $\ell=1 (2)$ only.
We see that $(63/4)^2\chi^{\rm F}_{n_{\rm f}}$ well scales with $\chi_{x^2+y^2+z^2}\propto\chi_{2z^2-x^2-y^2}$. 
This is due to the fact that the main contribution to Eq.~(\ref{eq:chi_E4_3}) comes from the 4f part $(\ell=1)$ and then Eq.~(\ref{eq:chi_E4_3}) is approximated 
as $\chi_{x^2+y^2+z^2}\approx |O^{x^2+y^2+z^2}_{\alpha 1\pm, \alpha 1\pm}|^2\chi^{\rm F}_{n_{\rm f}}$. 
Strictly speaking, the irreducible susceptibility which diverges at $T=0$ is $\chi^{\rm F}_{n_{\rm f}-n_{\rm d}}$, as noted in the eigenvalue analysis in Fig.3(b). However, $\chi^{\rm F}_{n_{\rm f}-n_{\rm d}}$ can be approximated as $\chi^{\rm F}_{n_{\rm f}-n_{\rm d}}\approx\chi^{\rm F}_{n_{\rm f}}$ in Eq.~(\ref{eq:chi_nf_nd}). 
Hence, $\chi_{x^2+y^2+z^2}$ shows enhancement for $T \to 0$, which is proportional to $\chi^{\rm F}_{n_{\rm f}}$. 

Interestingly, $\chi_{x^2-y^2}$ also increases for $T\to 0$ as shown in Fig.4(b), although the magnitude is three order smaller than $\chi_{x^2+y^2+z^2}$. 
We also plot the temperature dependence of $(3\sqrt{3}/5)^2\chi^{\rm F}_{n_{\rm d}}$ in Fig.4(b), which well scales with $\chi_{x^2-y^2}$. 
This implies that $\chi_{x^2-y^2}$ can be approximated as $\chi_{x^2-y^2}\approx|O^{x^2-y^2}_{\alpha 2\pm, \alpha 2\pm}|^2\chi^{\rm F}_{n_{\rm d}}$. 

These results indicate that from Eq.~(\ref{eq:elastic_C}) softening in elastic constants of not only the bulk modulus $C_{\rm B}\equiv (C_{11}+C_{22}+C_{33}+2C_{12}+2C_{13}+2C_{23})/9$ but also the shear moduli $C_{\rm u}\equiv (C_{11}+C_{22}+4C_{33}+2C_{12}-4C_{13}-4C_{23})/12$ and $C_{\rm v}\equiv (C_{11}+C_{22}-2C_{12})/4$ occur for low temperatures at the valence QCP.  
Here, $C_{\rm B}$, $C_{\rm u}$, and $C_{\rm v}$ are the elastic constants for the symmetry strains $\varepsilon_{xx}+\varepsilon_{yy}+\varepsilon_{zz}$, $2\varepsilon_{zz}-\varepsilon_{xx}-\varepsilon_{yy}$, and $\varepsilon_{xx}-\varepsilon_{yy}$, respectively. 
Namely, from the results shown in Figs.~\ref{fig:chi_E4}(a) and \ref{fig:chi_E4}(b) and Eq.~(\ref{eq:elastic_C}) the order of the magnitude of the elastic constants for $T\ll\Delta$ are estimated as $|C_{\rm B}| : |C_{\rm u}|: |C_{\rm v}|\approx 10^4 : 10^3 : 10$ when each quadrupole-strain coupling constant $g_{\Gamma}$ is assumed to be the same for $\Gamma=x^2+y^2+z^2$, $2z^2-x^2-y^2$, and $x^2-y^2$. 
The present study has revealed that if softening of $C_{\rm v}$ is observed 
 for $T\ll \Delta$
in $\beta$-YbAlB$_4$, it indicates that the mixture of the $J_z=\pm 1/2$ and $\mp 3/2$ states is realized in $|\Psi^{5{\rm d}}_{\rm \pm}\rangle$ as Eq.~(\ref{eq:5dWF}). 
This achieves the first direct observation of Yb 5d electron's contribution to the quantum critical state, which is of great significance. 

We also note that $\chi_{\Gamma}$ for $\Gamma=x^2, y^2$ and $z^2$ behave as $\chi_{x^2}\approx (19/4)^2\chi^{\rm F}_{n_{\rm f}}\sim\chi_{y^2}$ and $\chi_{z^2}\approx (25/4)^2\chi^{\rm F}_{n_{\rm f}}$, whose $T$ dependence can be seen by rescaling the data of $\chi^{\rm F}_{n_{\rm f}}$ in Fig.~\ref{fig:chi_E4}(a). 
This implies that softening of elastic constants of longitudinal modes $C_{11}$, $C_{22}$, and $C_{33}$ for strains $\varepsilon_{xx}$, $\varepsilon_{yy}$, and $\varepsilon_{zz}$ respectively occurs for $T\to 0$. 
No softening at least  for $T\to 0$ is expected in transverse modes $C_{44}$, $C_{55}$, and $C_{66}$ for strains $\varepsilon_{yz}$, $\varepsilon_{zx}$, and $\varepsilon_{xy}$ respectively because of vanishing of the form factors.

In $\alpha$-YbAl$_{1-x}$Fe$_x$B$_4$ with orthorhombic crystal structure (No.55 $Pbam$ $D_{2h}^{9}$), the Yb atom is also surrounded by 7 B rings as illustrated in Fig.~\ref{fig:Yb_B}(a)~\cite{Macaluso}. 
The sevenfold symmetry around Yb is broken by Al and/or Fe so that the mixture of the $J_z=\pm 1/2$ and $\mp 3/2$ states is expected in $|\Psi^{5{\rm d}}_{\pm}\rangle$ as Eq.~(\ref{eq:5dWF}). 
Furthermore, the Yb 4f CEF ground state as $a_{4{\rm f}}|\pm 5/2\rangle+b_{4{\rm f}}|\pm 1/2\rangle+c_{4{\rm f}}|\mp 3/2\rangle$ is suggested to be realized 
by the neutron~\cite{Broholm} and M\"{o}ssbauer~\cite{Kobayashi} measurements. 
In this case, the softening of $C_{\rm v}$ occurs more drastically since $\chi^{\rm F}_{n_{\rm f}}$ contributes to $\chi_{\rm v}$ in addition to $\chi^{\rm F}_{n_{\rm d}}$. 
It is interesting to observe the softening of $C_{\rm B}$, $C_{\rm u}$, and $C_{\rm v}$ for low temperatures in $\alpha$-YbAl$_{0.986}$Fe$_{0.014}$B$_4$.

The present RPA analysis has made it possible to identify which mode shows the softening in $C_{\Gamma}(T)$ for $T\to 0$. To clarify the temperature dependence of $C_{\Gamma}(T)$ accurately, the effect of the mode-mode coupling of critical Yb-valence fluctuations should be taken into account beyond the RPA. Such a calculation was performed in Ref.~\citen{WM2010} and for $\beta$-YbAlB$_4$ in Ref.~\citen{WM2014} where the valence susceptibility i.e., $\chi^{\rm F}_{n_{\rm f}-n_{\rm d}}$ is shown to behave as $\chi^{\rm F}_{n_{\rm f}-n_{\rm d}}\sim T^{-0.5}$ at the valence QCP. Hence, this temperature dependence is expected to appear in the elastic constants noted above.

\begin{acknowledgment}
The author acknowledges M. Yoshizawa who brought his attention to ultrasound measurements with enlightening discussions. 
He is grateful to R. Kurihara for valuable discussions about elastic constants. 
Thanks are also due to K. Miyake, Y. Kuramoto, H. Harima, C. Bareille, H. Kobayashi, and S. Wu for useful discussions. 
This work was supported by JSPS KAKENHI Grant Numbers JP18K03542, JP18H04326, and JP19H00648. 
\end{acknowledgment}


\end{document}